# New Bounds for Binary and Ternary Overloaded CDMA

Sh. Dashmiz
ACRI, Dept. of EE.
Sharif University of Technology
Tehran, Iran
shdashmiz@ee.sharif.edu

P. Pad
ACRI, Dept. of EE.
Sharif University of Technology
Tehran, Iran
pedram_pad@ee.sharif.edu

F. Marvasti
ACRI, Dept. of EE.
Sharif University of Technology
Tehran, Iran
marvasti@sharif.edu

*Abstract*—In this paper, we study binary and ternary matrices that are used for CDMA applications that are injective on binary or ternary user vectors. In other words, in the absence of additive noise, the interference of overloaded CDMA can be removed completely. Some new algorithms are proposed for constructing such matrices. Also, using an information theoretic approach, we conjecture the extent to which such CDMA matrix codes exist. For overloaded case, we also show that some of the codes derived from our algorithms perform better than the binary Welch Bound Equality codes; the decoding is ML but of low complexity.

## I. Introduction

Binary and ternary matrices are used in different fields of communications and coding theory. Binary matrices are the main material of Code Division Multiple Access (CDMA) systems. Ternary matrices are an important subject of study in some fields of coding theory. Matrices that introduce one-to-one mappings on the input vectors have special importance in these fields. Accordingly, four different scenarios can be deduced; the matrix is binary or ternary and the input vectors are binary or ternary. The maximum number of users (columns) for a fixed number of chip rate (rows) is the first natural question about these matrices. In what follows, we go over the literature and the applications of each of these four scenarios. As a benchmark, we compare our results for a spreading gain of 64.

*Scenario* 1: In binary inputs and synchronous overloaded CDMA systems, for errorless transmission under the assumption of noiseless channels, we have to use binary signature matrices coined as COW/COO[1] that introduce an injective transformation on all binary input vectors [1-3]. It is shown in [1-2] that if we fix the chip rate (number of rows) to 64, a COW/COO matrix with 164 users (columns) exists and there is no such a matrix with more than 268 columns. It is worth mentioning that the proof of the existence of a $64 \times 164$ COW/COO matrix is not constructive. Also, it was conjectured in [1-2] that a COW matrix with 239 columns exists. Additionally, we proposed an algorithm for fast implementation of the Maximum Likelihood (ML) decoder for a subclass of COW/COO matrices.

*Scenario* 2: In [3], a class of codes is suggested that has the ability of detection of active users for COW codes besides guaranteeing errorless transmission in a noiseless channel. These codes are named COWDA. In fact, they are $\{1, -1\}$-matrices that are injective on $\{1, 0, -1\}$-vectors. It is shown in [3] that with a fixed spreading gain of 64, the number of users is between 95 and 230. Again, the proof of existence of a $64 \times 95$ COWDA matrix is not constructive. A similar ML algorithm is proposed in [3] for decoding COWDA codes. Active user detection ability makes these codes appropriate for random access communication systems and ad-hoc networks [4-5].

*Scenario* 3: In [6-8] the problem of assigning binary code words to the binary values of each user is studied for unique decodability. This is equivalent to finding $\{1, 0, -1\}$-matrices that introduce a one-to-one mapping on $\{1, -1\}$-vectors. Moreover, these codes can be used in errorless synchronous CDMA systems. Employing ternary signatures instead of binary ones, we increase the number of users for a fixed chip rate. It is shown in [6-8] that a $64 \times 256$ $\{1, 0, -1\}$-matrix exists. Also, an upper bound for the number of users is derived in [6].

*Scenario* 4: The last case is to employ ternary matrices and ternary inputs. This case has not been reported in the literature. This scenario can be used for a CDMA system that employs ternary signatures and has the ability of errorless communication and detection of active users. As it is intuitively obvious, we will show that ternary signatures can handle larger number of users. These codes can be also decoded very fast using ML receivers.

In this paper, we study the above 4 scenarios using 3 steps. The first step is to propose a new method for constructing such matrices. The second step is to find an unachievable upper bound for the number of users for a given processing gain. The last step is to give an information theoretic conjecture for the maximum achievable number of users. For the first 3 scenarios, we claim that we have made some of the bounds reported in the literature tighter. Also, for the first scenario, we have extended our ML decoding of a $64 \times 104$ COW matrix to $64 \times 132$ COW matrix; its performance compared to single user and Welch Bound Equality (WBE) codes are also simulated. The results for the last scenario are completely new. Each of the scenarios is discussed in a sub-section in the following section. Section III consists of a brief summary of the results and some future works.

## II. Injective Binary/Ternary Matrices

This section consists of two main sub-sections. The first one is about binary matrices (Scenarios 1 and 2) and the second one is about ternary matrices (Scenarios 3 and 4). Each of these two sub-sections consists of two smaller sub-sections; the first one is about matrices which are injective on binary vectors (Scenario 1) and the second one is about matrices which are injective on ternary vectors (Scenario 2).

---
[1] Codes for Overloaded Wireless (COW) and Codes for Overloaded Optical (COO) Systems.

## II.1. BINARY MATRICES

In this section we study two classes of binary antipodal matrices that introduce one-to-one mappings on binary and ternary vectors. These matrices are named COW [1-2] and COWDA [3], respectively. Obviously, the second class is a sub-class of the first one. The following theorem is valid for both cases and proved in [1-2] and [3].

**Theorem 1** If $\mathbf{C}$ is a COW (COWDA) matrix and $\mathbf{P}$ is an invertible $\{1,-1\}$-matrix then $\mathbf{P} \otimes \mathbf{C}$ is a COW (COWDA) matrix, where $\otimes$ denotes the Kronecker product.

### II.1.1 BINARY MATRIX, BINARY INPUT VECTORS

In this sub-section we search for $\{1,-1\}$-matrices that introduce one-to-one mappings on $\{1,-1\}$-vectors, i.e., the so called COW codes used in wireless CDMA [1-2]. It was proved in [1-2] that if an $m \times n$ COW matrix exists, then a $2m \times (2n + \lceil (m-1)\log_3 2 \rceil) \approx 2m \times (2n + 0.63m)$ COW matrix exists. The weakness of this theorem is that it only proves the existence of such matrices and does not propose a constructive method. As a special case, this theorem infers that a $64 \times 164$ COW matrix exists. The main contribution of this section is to prove a theorem which introduces a $64 \times 193$ COW matrix as a special case.

**Theorem 2** Assume $\mathbf{C}$ is an $m \times n$ COW matrix, $\mathbf{H}_2 = \begin{bmatrix} +1 & +1 \\ +1 & -1 \end{bmatrix}$ and $\mathbf{V}_{2m \times (m-1)} = [v_{ij}]$ such that

$$v_{ij} = \begin{cases} -1 & j < i \leq m \\ +1 & \text{otherwise} \end{cases}.$$

$\mathbf{D} = [\mathbf{H}_2 \otimes \mathbf{C} \ \mathbf{V}]$ is an $2m \times (2n + m - 1)$ COW matrix.

For the proof, please refer to Appendix A.

Starting from $\mathbf{H}_2$ (which is trivially a COW matrix) and using Theorem 2 recursively, we arrive at $\mathbf{C}_{4 \times 5}$, $\mathbf{C}_{8 \times 13}$, $\mathbf{C}_{16 \times 33}$, $\mathbf{C}_{32 \times 81}$ and $\mathbf{C}_{64 \times 193}$ COW matrices.

Define the channel capacity of a CDMA system as [1-2]:

$$C_I(m,n) = \max_{\mathbf{A} \in \mathcal{M}_{m \times n}(\{1,-1\})} \log_2 |\mathbf{A}\{1,-1\}^n|$$

where $|\ |$ denotes the cardinality of a set and $\mathcal{M}_{m \times n}(\{1,-1\})$ is the set of all $m \times n$ $\{1,-1\}$-matrices. Note that if $C_I(m,n) = n$, this means that there exists an $m \times n$ COW matrix. It is proved in [1-2] that

$$C_I(m,n) \geq n - \log_2 \sum_{j=0}^{\lfloor \frac{n}{2} \rfloor} \binom{n}{2j} \left( 2^{-2j} \binom{2j}{j} \right)^m \quad (1)$$

The right-hand side of the above equation is plotted in Fig. 1. Using this curve, authors of [1-2] conjectured that there exists a $64 \times 239$ COW matrix (i.e., at most 239 users). In summary, we state that for a fixed number of chip rate 64, we can have a COW matrix of size $64 \times 193$. From [2], there is no COW matrix with more than 268 columns and thus we conjecture the existence of a COW matrix with 239 columns.

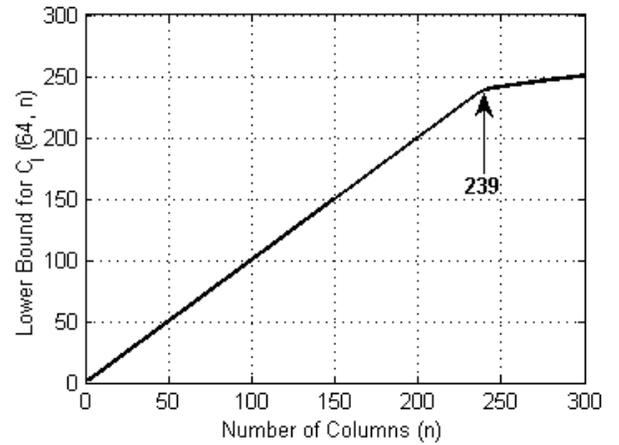

Fig. 1. A lower bound for $C_I(m,n)$ versus $n$ when $m = 64$

Because COW codes are suitable for overloaded binary synchronous CDMA systems, we are interested in their performances in a noisy environment. We have simulated the performance of $\mathbf{C}_{64 \times 132} = \mathbf{H}_4 \otimes \mathbf{C}_{16 \times 33}$ which is a COW matrix (according to Theorem 1) in an AWGN channel. We use the simple ML decoder which is proposed in [2]. These codes are also compared with Binary WBE sequences in conjunction with iterative decoding. The results are depicted in Fig. 1. It is very interesting that despite a high percentage of the overloading factor, these codes show a better performance. Thus, applying them in practical situations can handle many more users (more than twice of that of Walsh-Hadamard codes) without increasing the bandwidth.

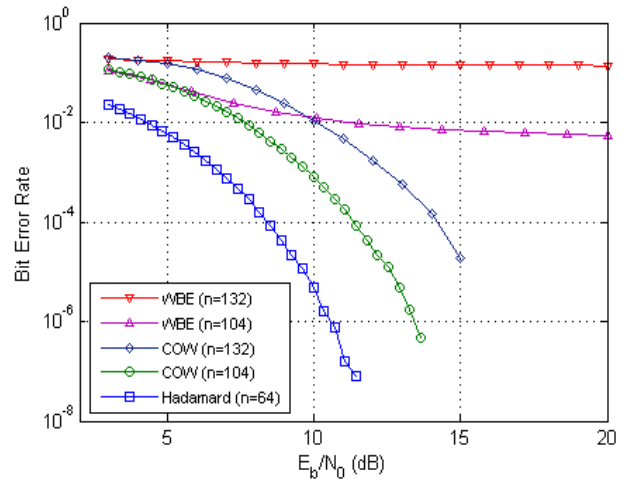

Fig. 2. Comparison of Hadamard (single user), COW, and WBE codes in an AWGN channel; $m = 64$.

### II.1.2 BINARY MATRIX, TERNARY VECTORS

In this part we wish to modify Theorem 2 and derive a similar theorem when we are looking for $\{1,-1\}$-matrices that are injective on $\{1,0,-1\}$-vectors. These codes are the ones that have to be used when we need a CDMA system that not only is errorless but also has the property of detection of active users. These matrices are named COWDA in [3]. According to [3], if an $m \times n$ COWDA matrix exists, then a $4m \times (4n + \lceil (m-2)\log_5 2 \rceil) \approx 4m \times (4n + 0.43m)$ COWDA matrix exists. This theorem again discusses the existence of such matrices and is not constructive. As a special case, this theorem guarantees the existence of a $64 \times 95$ COWDA matrix. In the following we propose a theorem that constructs a $64 \times 103$ COWDA matrix.

**Theorem 3** Assume **C** is an $m \times n$ COWDA matrix, $\mathbf{H}_4 = \mathbf{H}_2 \otimes \mathbf{H}_2$ and $\mathbf{V}_{4m \times (m-1)} = [v_{ij}]$ such that

$$v_{ij} = \begin{cases} -1 & j < i \leq m \\ +1 & \text{otherwise} \end{cases}.$$

Then, $\mathbf{D} = [\mathbf{H}_4 \otimes \mathbf{C} \ \mathbf{V}]$ is a $4m \times (4n + m - 1)$ COWDA matrix.

The proof is in Appendix B.

As a special case, if we start with the COWDA matrix $\mathbf{C}_{16 \times 22}$ proposed in [3], applying Theorem 3, we obtain a $64 \times 103$ COWDA matrix.

Now, define the channel capacity as before
$$C_{II}(m,n) = \max_{\mathbf{A} \in \mathcal{M}_{m \times n}(\{1,-1\})} \log_3 |\mathbf{A}\{1,0,-1\}^n|.$$

Likewise, using arguments similar to (1), we have
$$C_{II}(m,n) \geq 2n - \log_3 A(m,n) \quad (2)$$

where

$$A(m,n) = \sum_{a=0}^{n} 2^a \binom{n}{a} \sum_{\substack{x \leq a \\ y \leq n-x}} B(x,y) \binom{a}{x}$$
$$\cdot \left(2^{-(x+y)} \sum_{r} \binom{x}{r}\binom{y}{x - 2r + \frac{y}{2}}\right)^m$$

and

$$B(x,y) = \sum_{t=0}^{y} 2^t \binom{a-x}{y-t}\binom{n-a}{t}.$$

The right-hand side of (2) is plotted in Fig. 3.

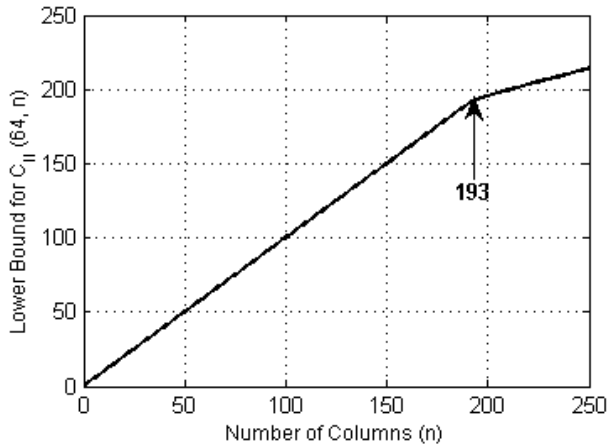

Fig. 3. A lower bound for $C_{II}(m,n)$ versus $n$ when $m = 64$.

In summary, in this section we derived that for a fixed chip rate equal to 64, there exists a COWDA matrix with 103 users. Also, from Fig. 3, we conjecture that a $64 \times 193$ COWDA matrix exists. We also know from [3] that there is no COWDA matrix with more than 230 columns.

### II.2. TERNARY MATRICES

In this section we will discuss two classes of Ternary matrices that introduce one-to-one transformations on binary and ternary vectors. The first class was studied in [6-8] as T-codes but the second one has not been known before. We coin this class as TCOW and TCOWDA, respectively that are a sub-class of T-codes. The following theorem is valid for both classes and can be easily proved in a similar way as Theorem 1.

**Theorem 4** If **C** is a TCOW (TCOWDA) matrix and **P** is an invertible $\{1,0,-1\}$-matrix, then $\mathbf{P} \otimes \mathbf{C}$ is a TCOW (TCOWDA) matrix.

In the following subsections, we study these two classes separately.

#### II.2.1 TERNARY MATRIX, BINARY VECTORS

The binary T-code proposed in [6] can be converted into the problem of finding $\{1,0,-1\}$-matrices that are injective on $\{1,-1\}$-input vectors. We can use these codes for errorless CDMA systems with ternary signatures (instead of COW matrices which are binary). Using ternary signatures, we can handle many more number of users. We name these matrices as TCOW matrices. It is shown in [6-8] that with a fixed number of chips of 64, there exists a TCOW matrix with 256 users; also, they have shown that there is no TCOW matrix with more than 330 users. Here, we first propose a theorem similar to Theorem 2 about TCOW matrices. Although the new theorem does not extend the number of users more than that of [6-8], it gives a new constructive method for TCOW matrices. However, the main contribution of this section is to derive a formula similar to (1) and (2) that we can conjecture about the maximum size of TCOW matrices.

**Theorem 5** Assume **C** is an $m \times n$ COW matrix and $\mathbf{V}_{m \times (m-1)} = [v_{ij}]$ such that

$$v_{ij} = \begin{cases} 0 & j < i \leq m \\ 1 & \text{otherwise} \end{cases}.$$

Then, $\mathbf{D} = [\mathbf{C} \ \mathbf{V}]$ is a $m \times (n + m - 1)$ TCOW matrix.

Applying this theorem on the COW matrix $\mathbf{C}_{64 \times 193}$, we can obtain a $64 \times 256$ TCOW matrix.

The proof is very similar to the proves of Theorems 2 and 3.

Now, define
$$C_{III}(m,n) = \max_{\mathbf{A} \in \mathcal{M}_{m \times n}(\{1,0,-1\})} \log_2 |\mathbf{A}\{1,-1\}^n|.$$

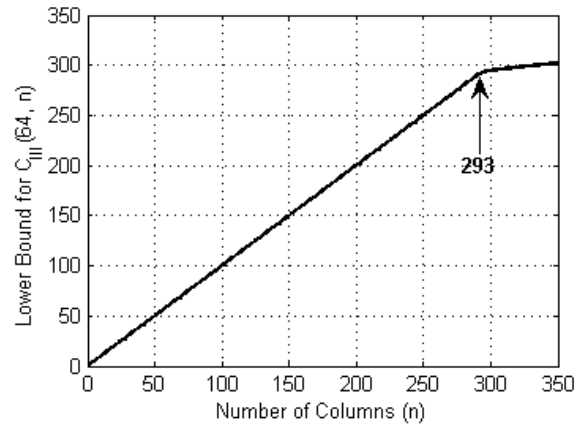

Fig. 4. The lower bound (3) for $C_{III}(m,n)$ versus $n$ when $m = 64$.

Similar to (1) and (2) it is easy to show that

$$C_{III}(m,n) \geq n - \log_2 \sum_{k=0}^{n} \binom{n}{k} \left( 3^{-k} \sum_{j=0}^{\lfloor \frac{k}{2} \rfloor} \binom{k}{2j}\binom{2j}{j} \right)^m \quad (3)$$

Again, the right-hand side of (3) is depicted in Fig. 4 when $m = 64$.

On the whole, according to [6-8] we can say that for a fixed number of chip rates (rows) of 64, we have a TCOW matrix with 256 users (columns) and there is no such a matrix with more than 330 columns. Moreover, from Fig. 4 we can conjecture that there are TCOW matrices with 293 users when $m$ is fixed to 64.

### II.2.1 TERNARY MATRIX, TERNARY VECTORS

These matrices have not been studied in the literature and will be covered here. Similar to the previous cases, these codes can be used for errorless CDMA systems with the ability of active user detection. Using ternary signatures has the advantage that although the hardware is not very complicated, the system capacity increases and can handle many more users. These matrices are called TCOWDA matrices. In the following, we propose a method for constructing such codes, develop an upper bound for the number of users for a given number of chip rates and conjecture about the maximum size of a TCOWDA matrix.

**Theorem 6** Assume $\mathbf{C}$ is an $m \times n$ COWDA matrix and $\mathbf{V}_{2m \times (m-1)} = [v_{ij}]$ such that

$$v_{ij} = \begin{cases} 0 & j < i \leq m \\ 1 & \text{otherwise} \end{cases}.$$

$\mathbf{D} = [\mathbf{H}_2 \otimes \mathbf{C} \;\; \mathbf{V}]$ is a $2m \times (2n + m - 1)$ TCOWDA matrix.

The proof is very similar to the proves of Theorems 2 and 3 and are eliminated due to the 5 page limitation.

Applying Theorem 1 on a $16 \times 22$ COWDA matrix, we can obtain a $32 \times 44$ COWDA matrix. Now, applying Theorem 6 on this COWDA matrix, we find a $64 \times 119$ TCOWDA matrix.

Following the method used in [3] for obtaining an upper bound for the number of uerrs of a COWDA matrix, we can derive an upper bound for the number of uerrs of a TCOWDA matrix. But noting that the entropy of a single row of a TCOWDA matrix is maximized when all of its entries are non-zero, the derived upper bound is the same as that of [3]. Thus, in the special case where the spreading gain is 64, 230 is an upper bound for the number of users similar to the COWDA case.

Now, define the capacity as before
$$C_{IV}(m,n) = \max_{\mathbf{A} \in \mathcal{M}_{m \times n}(\{1,0,-1\})} \log_3 |\mathbf{A}\{1, 0, -1\}^n|.$$

Similar to (1), (2) and (3), we can show that
$$C_{IV}(m,n) \geq 2n - \log_3 A(m,n) \quad (4)$$
where
$$A(m,n) = \sum_{a=0}^{n} 2^a \binom{n}{a} \sum_{\substack{x \leq a \\ y \leq n-x}} B(x,y) \binom{a}{x}$$
$$\cdot \left( 3^{-(x+y)} \sum_{q,w,r} \binom{x}{q}\binom{y}{w}\binom{x-q}{r}\binom{y-w}{x-q-2r+\frac{y-w}{2}} \right)^m$$

and
$$B(x,y) = \sum_{t=0}^{y} 2^t \binom{a-x}{y-t}\binom{n-a}{t}.$$

However, this formula does not have a sharp knee point like what can be seen in Figs. 1, 3 and 4. Thus, no feasible conjecture about the maximum number of users for a TCOWDA matrix of fixed $m$ can be made from using this formula. For this reason, we use another method for conjecture the maximum number of users of a TCOWDA matrix when $m = 64$.

Using (2), we find that there is a knee point at $n = 81$ when $m = 32$. Now, according to Theorem 6, if a $32 \times 81$ COWDA matrix exists, then a $64 \times 193$ TCOWDA matrix also exists. It is our conjecture that 193 is the maximum number of users for TCOWDA matrix.

In summary, for a fixed number of chips 64, we have a TCOWDA matrix with 119 columns, and we know that there is no matrix with 230 columns. Thus, we conjecture the existence of a matrix with 193 users.

FInally, we can use the same method that was used for conjecturing the maximum number of users of a TCOWDA matrix for the COWDA for TCOW and COW matrices. According to Theorem 5 if a $64 \times n$ COW matrix exits, then a $64 \times (n + 63)$ TCOWDA matrix also exists. Referring to Figs. 1 and 3, we can conjecture an interval for the maximum number ofusers of a COW or TCOW matrices. For COW matrices, the maximum number of columns is between $230 (= 293 - 63)$ and 239. And for TCOW matrices, this number is between 293 and $302 (= 239 + 63)$.

### III. CONCLUSION

In this paper, we have studied binary and ternary matrices that introduce one-to-one mappings on binary or ternary input vectors. All four scenarios have been studied and simulated.

The matrices that are injective on binary vectors (COW and TCOW) are suitable for synchronous CDMA systems. The performance of a highly overloaded CDMA system (with 64 chips and 132 users) that employs COW codes in an AWGN is depicted in Fig. 2. Comparing this curve with the WBE codes and the best of current decoding algorithms showed a significant improvement.

The matrices that are injective on ternary vectors (COWDA and TCOWDA) are suitable for CDMA systems that have the errorless property as well as active user detection ability. The performance of COWDA codes in noisy channel is discussed in [3]. Using TCOWDA matrices, we increase the number of users without losing the errorless transmission.

All our results together with those of [1-3] and [6-8] are depicted in Table 1. The bold parts are related to the contribution of this paper. Note that, using the theorems of this paper, we can construct matrices that can achieve the lower bounds.

In the method used for deriving the upper bounds, using the joint entropy of every two or more entries of the coded vectors gives tighter upper bounds that will be considered in future.

Also, all the above results can be extended to the general case of $m$-ary matrices which are injective on $n$-ary vectors.

| Result Type | Lower Bound | Upper Bound | Conjecture |
|---|---|---|---|
| COW | **193** | 268 [2] | **230~239** |
| COWDA | **103** | 230 [3] | **193** |
| TCOW | 256 [6-8] | 330 [6] | **293~302** |
| TCOWDA | **119** | 230 | **193** |

Table1. Maximum number of columns (users) for a matrix of 4 different scenarios with 64 rows (chip rate).

## Appendix A: Proof of Theorem 2

Suppose $\mathbf{V} = \begin{bmatrix} \mathbf{M}_1 \\ \mathbf{M}_2 \end{bmatrix}$ where $\mathbf{M}_1$, $\mathbf{M}_2$ are $m \times (m-1)$ matrices. Suppose there exists a $\{1,0,-1\}$-vector $Z = \begin{bmatrix} Z_1 \\ Z_2 \end{bmatrix}$ such that $\mathbf{D}Z = 0$ and $Z_1$ and $Z_2$ are $2n \times 1$ and $(m-1) \times 1$ vectors, respectively. If $Z_1 = \begin{bmatrix} Z_3 \\ Z_4 \end{bmatrix}$ with $Z_3, Z_4 \in \{1,0,-1\}^n$, we have

$$(\mathbf{H}_2 \otimes \mathbf{C})Z_1 + \mathbf{V}Z_2 = 0.$$

Multiplying both sides by $\frac{1}{2}\mathbf{H}_2 \otimes \mathbf{I}_m$, we arrive at

$$(\mathbf{I}_2 \otimes \mathbf{C})Z_1 = \frac{-1}{2}(\mathbf{H}_2 \otimes \mathbf{I}_m)\mathbf{V}Z_2$$

which implies

$$\begin{bmatrix} \mathbf{C} & 0 \\ 0 & \mathbf{C} \end{bmatrix} Z_1 = \begin{bmatrix} \mathbf{I}_m & \mathbf{I}_m \\ \mathbf{I}_m & -\mathbf{I}_m \end{bmatrix} \begin{bmatrix} \frac{-\mathbf{M}_1}{2} Z_2 \\ \frac{-\mathbf{M}_2}{2} Z_2 \end{bmatrix} = \begin{bmatrix} \frac{-(\mathbf{M}_1+\mathbf{M}_2)}{2} Z_2 \\ \frac{(\mathbf{M}_2 - \mathbf{M}_1)}{2} Z_2 \end{bmatrix}.$$

Decoupling the equations, we arrive at

$$\mathbf{C}Z_3 = \frac{-(\mathbf{M}_1+\mathbf{M}_2)}{2} Z_2 \quad (A1)$$

$$\mathbf{C}Z_4 = \frac{(\mathbf{M}_2-\mathbf{M}_1)}{2} Z_2 \quad (A2)$$

Because $\mathbf{C}$ is a $\{1,-1\}$-matrix and $Z_3$ is a $\{1,0,-1\}$-vector, $\frac{1}{2}(\mathbf{M}_1+\mathbf{M}_2)Z_2$ is a vector that has the same parity in all entries. Now, note that

$$\frac{(\mathbf{M}_1+\mathbf{M}_2)}{2} = \begin{bmatrix} 1 & \cdots & 1 \\ 0 & \ddots & \vdots \\ \vdots & \ddots & 1 \\ 0 & \cdots & 0 \end{bmatrix}.$$

Thus, if $z_{i_1}, z_{i_2}, \ldots, z_{i_k}$ for $i_1 < i_2 < \cdots < i_k$ are all non-zero entries of $Z_2$ and $\frac{1}{2}(\mathbf{M}_1+\mathbf{M}_2)Z_2 = [w_1 \cdots w_m]^T$, it is obvious that $w_{i_1} \equiv k$ and $w_{i_1+1} \equiv k-1 \pmod{2}$ and this contradicts with the above mentioned fact. Thus $Z_2 = 0$ and from $(A1)$ and $(A2)$ we conclude that $\mathbf{C}Z_3 = \mathbf{C}Z_4 = 0$. Therefore, because $\mathbf{C}$ is a COW matrix, we have $Z_3 = Z_4 = 0$ and consequently $Z_1 = 0$ and $Z = 0$. And this completes the proof.

## Appendix B: Proof of Theorem 3

Let $\mathbf{V} = [\mathbf{M}_1^T \ \mathbf{M}_2^T \ \mathbf{M}_3^T \ \mathbf{M}_4^T]^T$ where $\mathbf{M}_i$ is an $m \times (m-1)$ matrix, for $i = 1, \ldots, 4$. Suppose there exist a vector $Z = \begin{bmatrix} Z_1 \\ Z_2 \end{bmatrix}$ with $Z_1 \in \{\pm 2, \pm 1, 0\}^{4n}$ and $Z_2 \in \{\pm 2, \pm 1, 0\}^{m-1}$ such that $\mathbf{D}Z = 0$. If $Z_1 = [W_1^T \ W_2^T \ W_3^T \ W_4^T]^T$ with $W_i \in \{\pm 2, \pm 1, 0\}^n$, for $i = 1, \ldots, 4$, we have

$$(\mathbf{H}_4 \otimes \mathbf{C})Z_1 + \mathbf{V}Z_2 = 0$$

Multiplying both sides by $\frac{1}{4}\mathbf{H}_4 \otimes \mathbf{I}_m$, we arrive at

$$(\mathbf{I}_4 \otimes \mathbf{C})Z_1 = \frac{-1}{4}(\mathbf{H}_4 \otimes \mathbf{I}_m)\mathbf{V}Z_2.$$

Thus

$$\mathbf{C}W_1 = \frac{-(\mathbf{M}_1+\mathbf{M}_2+\mathbf{M}_3+\mathbf{M}_4)}{4} Z_2 \quad (B1)$$

Because $\mathbf{C}$ is a COWDA matrix and $W_1$ is a $\{\pm 2, \pm 1, 0\}$-vector, $\frac{1}{4}(\mathbf{M}_1+\mathbf{M}_2+\mathbf{M}_3+\mathbf{M}_4)Z_2$ is a vector with all entries have the same parity.

Now, note that

$$\frac{(\mathbf{M}_1+\mathbf{M}_2+\mathbf{M}_3+\mathbf{M}_4)}{4} = \begin{bmatrix} 1 & \cdots & 1 \\ \frac{1}{2} & \ddots & \vdots \\ \vdots & \ddots & 1 \\ \frac{1}{2} & \cdots & \frac{1}{2} \end{bmatrix}.$$

Let $z_{i_1}, z_{i_2}, \ldots, z_{i_k}$ for $i_1 < i_2 < \cdots < i_k$ bee all non-zero entries of $Z_2$ and $\frac{1}{4}(\mathbf{M}_1+\mathbf{M}_2+\mathbf{M}_3+\mathbf{M}_4)Z_2 = [w_1 \cdots w_m]^T$. It is obvious that if $z_{i_1} = \pm 2$, then $p_{i_1} \equiv p_{i_1+1} + 1 \pmod{2}$ and if $z_{i_1} = \pm 1$, then $p_{i_1+1} = p_{i_1} + t + \frac{1}{2}$ where $t$ is integer, which contradicts with $(B1)$ because the left-hand side of $(B1)$ is an integer vector and only one of the $p_{i_1}$ or $p_{i_1+1}$ can be integer. So $Z_2 = 0$ and we conclude that $\mathbf{C}W_1 = \mathbf{C}W_2 = \mathbf{C}W_3 = \mathbf{C}W_4 = 0$. And since $\mathbf{C}$ is a COWDA matrix, we have $W_1 = W_2 = W_3 = W_4 = 0$, $Z_1 = 0$ and $Z = 0$ which completes the proof.